# Towards Sustainable IoT: Challenges, Solutions, and Future Directions for Device Longevity

Ghazaleh Shirvani
ghazalehshirvani@cmail.carleton.ca
Carleton University
Ottawa, ON, Canada

Saeid Ghasemshirazi
saeidghasemshirazi@cmail.carleton.ca
Carleton University
Ottawa, ON, Canada

## ABSTRACT

In an era dominated by the Internet of Things, ensuring the longevity and sustainability of IoT devices has emerged as a pressing concern. This study explores the various complex difficulties which contributed to the early decommissioning of IoT devices and suggests methods to improve their lifespan management. By examining factors such as security vulnerabilities, user awareness gaps, and the influence of fashion-driven technology trends, the paper underscores the need for legislative interventions, consumer education, and industry accountability. Additionally, it explores innovative approaches to improving IoT longevity, including the integration of sustainability considerations into architectural design through requirements engineering methodologies. Furthermore, the paper discusses the potential of distributed ledger technology, or blockchain, to promote transparent and decentralized processes for device provisioning and tracking. This study promotes a sustainable IoT ecosystem by integrating technology innovation, legal change, and social awareness to reduce environmental impact and enhance resilience for the digital future.

## 1 INTRODUCTION

The increasing spread of Internet of Things devices across different sectors signals a new era of connectivity and technical progress. These technologies, which include smart household appliances and industrial sensors, provide unparalleled levels of ease, efficiency, and automation. Yet, a significant issue within this rapid growth is the durability of these devices. IoT devices function under ever-changing and challenging conditions, which might affect how long they remain operational, unlike conventional software systems.

The software industry has dealt with lifespan challenges through procedures like as software evolution and maintenance, but the situation is different for IoT devices. These devices are affected by software upgrades, hardware deterioration, ambient conditions, and changing connectivity standards. To ensure the durability and lifetime of IoT devices, a comprehensive understanding and a proactive strategy are required to tackle these complex difficulties.

This introduction aims to highlight the importance of digital longevity in the IoT field, outlining essential concepts and stressing the need for interdisciplinary efforts to prioritize longevity in the design of IoT devices from the beginning[11]. We will investigate the various obstacles that affect the lifespan of IoT devices, such as user misunderstandings, consumer behavior influenced by trends, and hardware restrictions. Our goal is to emphasize the importance of creative design approaches and educating consumers.

The following parts of this article will explore a detailed classification of factors causing the early decommissioning of IoT devices, with the goal of reducing the electronic waste produced by outdated IoT equipment. Our goal is to analyze root causes and systemic obstacles to provide practical solutions that can reduce environmental effects and improve the durability of IoT devices. By collaborating across disciplines and making focused efforts to tackle these difficulties, we can provide a basis for a more sustainable and resilient Internet of Things ecosystem.

## 2 BACKGROUND
### 2.1 IoT Target Definition

The IoT consists of a wide range of networked devices that are present in different parts of modern life, including wearable electronics such as smartwatches and advanced sensors found in autonomous vehicles. Yet, a significant pattern within this vast environment is the early retirement of IoT devices, especially those that are inexpensive and easily available to regular users. Although common, many users have a restricted comprehension of the fundamental structure and operation of these gadgets. Therefore, when encountering any perceived issue, people frequently choose to buy a new replacement instead of investigating viable troubleshooting or repair solutions[17]. Consumer behavior is impacted by aesthetic preferences and the appeal of newer device models, causing them to discard fully functional IoT gadgets just due to a lack of the latest design features or visual attractiveness. It is important to understand that this phenomena goes beyond just device aging, since new technology features and design aesthetics make older devices appear outdated to users[20].

This study focuses on exploring the problems related to the durability of small-scale IoT devices that are affordable and have minimal impact inside larger IoT networks. Although essential to consumers' everyday activities, these devices may not receive substantial focus in the larger IoT ecosystem that includes networked smart infrastructure and industrial systems. The combined influence of these factors on consumer behavior, resource use, and electronic waste production highlights the need of dealing with problems associated their early decommissioning[9].

This study tries to identify the elements that cause premature obsolescence in IoT devices and suggest specific solutions to promote a more sustainable and responsible approach to using and disposing of these devices. We want to prolong the operational lives of IoT devices and reduce their environmental impact by educating users, implementing innovative designs, and following industry standards. This is in line with the ideas of circular economy and sustainable development.

## 2.2 Related Studies

The widespread adoption of Internet of Things across diverse sectors underscores their critical role in modern society. However, ensuring their long-term sustainability presents a significant challenge. While historical efforts have primarily focused on extending the lifespan of software systems through evolution and maintenance, the complex factors influencing the durability of IoT devices have often been overlooked[1].

Studies show that most customers lack a deep grasp of IoT devices, which causes them to replace gadgets prematurely if they are seen as obsolete or unsafe. Moreover, consumer behavior driven by trends worsens this problem, as people prioritize superficial characteristics above long-term practicality and sustainability[12]. Intrinsic hardware restrictions, such as processing power and energy sources, speed up the obsolescence of IoT devices. This requires creative design strategies to improve upgradeability and flexibility.

Recent studies highlight the impact of physical degradation mechanisms, including corrosion and component fatigue, on the operational sustainability of IoT devices[18]. Proactive maintenance procedures and component-level monitoring are proposed solutions to mitigate the effects of environmental conditions and usage patterns on device longevity. Leveraging advanced technologies such as predictive maintenance algorithms and self-healing systems, the IoT industry aims to enhance the durability and longevity of devices across various operational settings.

Collaboration between software and hardware experts is crucial for longevity of IoT devices. Innovative design methods, together with consumer education programs and preventive maintenance plans, are crucial elements of a comprehensive strategy to establish a sustainable and robust IoT environment. Collaboration among industry stakeholders may reduce environmental effects and enhance the durability of IoT devices, therefore promoting the advancement of linked technologies.[6]

## 2.3 Challenges and Opportunities in IoT Longevity

Digital longevity refers to the ability of information systems to remain functional, relevant, and usable over long periods of time, adapting to changing socio-technological environments. This idea emphasizes the complex and ever-changing nature of longevity in the context of IoT devices. At the core of this comprehension is the acknowledgment of the complex interaction among different elements, such as information objects, system and software components, and the overarching information systems that coordinate their functions[14].

An important obstacle in maintaining the durability of IoT devices is managing the intricate set of elements that impact their lifespan. These elements go beyond only hardware and software considerations to include wider societal, economic, and environmental issues[3]. Technological obsolescence, caused by fast developments in hardware capabilities and software functionality, is a major threat to the continued relevance of IoT devices. Furthermore, changing legal frameworks and standards, together with developing customer preferences and commercial demands, add to the difficulties related to the lifetime of IoT.

Nevertheless, within these constraints are numerous chances for creativity and progress. By taking a comprehensive strategy to tackle longevity issues, participants in the IoT ecosystem can discover new opportunities for sustainable growth and societal influence. This involves improving the longevity and ability to be upgraded of hardware and software components, as well as increasing knowledge and involvement among end-users in understanding the importance of durability and responsible consumption. Furthermore, the combination of new technologies like artificial intelligence, edge computing, and block chain presents great opportunities to enhance the durability of IoT devices. These solutions improve the robustness and adaptability of IoT ecosystems by enabling proactive maintenance, predictive analytics, and decentralized governance mechanisms to address growing difficulties.

## 2.4 Research Questions

In order to provide guidance for the advancement of more resilient Internet of Things ecosystems, the present study aims to investigate the subsequent research questions:

**RQ1.** What are the primary factors contributing to the premature decommissioning of IoT devices among end-users, and how can these challenges be mitigated through informed design and consumer engagement strategies?

**RQ2.** How do changing social and technological environments impact the durability of IoT devices, and how do regulations, market forces, and consumer actions affect the lifespan of these devices?

**RQ3.** What innovative design paradigms and technological solutions can be leveraged to enhance the durability, upgradability, and adaptability of IoT hardware and software components, thereby extending their functional lifespan?

**RQ4.** How can interdisciplinary approaches, drawing insights from fields such as digital curation, sustainability science, and human-computer interaction, inform the development of more resilient and sustainable IoT ecosystems?

**RQ5.** To what extent do emerging technologies, including artificial intelligence, edge computing, and blockchain, hold promise for addressing challenges related to IoT device longevity, and what are the key considerations in integrating these technologies into existing IoT infrastructure?

## 3 EARLY IOT DECOMMISSIONING FACTORS TAXONOMY

While academic resources specifically addressing the longevity of IoT devices are limited, insights from studies on computing systems longevity and resilience, as well as related IoT lifespan literature[5], have been instrumental in constructing a taxonomy of impacting factors. Through a comprehensive review of the literature, several primary factors have been identified, including security, privacy, user knowledge gaps, inadequate legislation, and vendor agility. Each of these factors plays a significant role in influencing the premature decommissioning of IoT devices, and a deeper exploration of each is warranted. The summary of challanges and solutions are illustrated in Figure 1.

### 3.1 User Awareness Gap

One of the most critical factors contributing to premature IoT decommissioning is the lack of user knowledge regarding these devices. A significant reason leading to early IoT decommissioning is users' lack of information about these devices. Research regularly demonstrates that numerous users do not possess a thorough comprehension of IoT devices, their functions, and troubleshooting techniques. Studies conducted by [24] and [10] have shown that less educated consumers frequently have difficulty identifying problems or faults in their IoT devices, as revealed by surveys. Therefore, when encountering perceived issues, consumers with inadequate knowledge may choose to replace their equipment instead of trying to fix the problem.

This situation greatly adds to the accumulation of electronic trash in the Internet of Things sector, especially when combined with advertising practices that highlight the appeal of newer gadgets with apparently improved functionality. Some consumers may not be aware of the full capabilities of their existing gadgets and may mistakenly believe that they do not have certain features that are promoted in newer versions. Additionally, their lack of knowledge about existing updates and upgrades worsens the situation, leading them to potentially dispose of equipment that may be repaired or enhanced with software updates or maintenance.

**Challenges:**

- Troubleshooting Difficulties: Users may lack the skills to identify and resolve issues with their IoT devices, leading to frustration and premature replacement.
- Limited Understanding of Features: Many consumers may not fully comprehend the capabilities of their devices, resulting in the premature discarding of fully functional IoT gadgets.

**Solutions:**

- User Education Programs: Implementing comprehensive user education programs can empower consumers with the knowledge needed to troubleshoot and maintain their devices effectively.
- Enhanced User Manuals: Improving the clarity and accessibility of user manuals can bridge the knowledge gap, providing users with valuable insights into the features and potential issues of their IoT devices.

### 3.2 Security Vulnerabilities

Enhancing the security and privacy features of IoT devices is crucial for guaranteeing their long-term viability and endurance. Researchers widely agree that strong security measures integrated in the design process may greatly reduce IoT waste and prolong the lifespan of devices. Integrating these measurements encounters several problems that must be resolved to advance sustainable IoT activities[19].

One problem is motivating manufacturers to emphasize security in their design processes. Manufacturers may emphasize cost and time-to-market above security, even though effective security measures are crucial, particularly when immediate advantages are not clear. This underscores the necessity of regulatory frameworks or industry standards that motivate manufacturers to include security measures into their goods.

It is essential for the long-term viability of IoT devices to have adaptable security mechanisms that can respond to changing threats. Technological progress requires security systems that can endure new dangers as they arise. Continuous research and development are needed to establish flexible security frameworks that can be regularly updated and sustained during the device's lifespan[22].

Moreover, the expenses and complexity associated with incorporating extensive security protocols might hinder general acceptance. IoT devices equipped with sophisticated security measures might have a higher production cost, which would restrict their availability to customers. Ensuring that safe IoT solutions are available to a wide variety of users requires balancing security needs with price and usability[7].

Addressing these challenges requires a multidisciplinary approach that combines technological innovation with regulatory intervention and industry collaboration. To achieve a more safe and sustainable future for the IoT ecosystem, manufacturers should be encouraged to emphasize security, innovation in security solutions should be promoted, and consumer awareness should be increased.

**Challenges:**

- Lack of Manufacturer Emphasis on Security: Manufacturers may prioritize cost and time-to-market over robust security measures, leaving devices vulnerable to evolving cyber threats.
- Adaptability to Emerging Threats: IoT devices need flexible security mechanisms that can adapt to new and evolving cybersecurity risks. Failure to do so can render devices obsolete and insecure.

**Solutions:**

- Regulatory Frameworks: Implementing regulatory frameworks that mandate manufacturers to prioritize security in their design processes can ensure a baseline level of security in IoT devices.

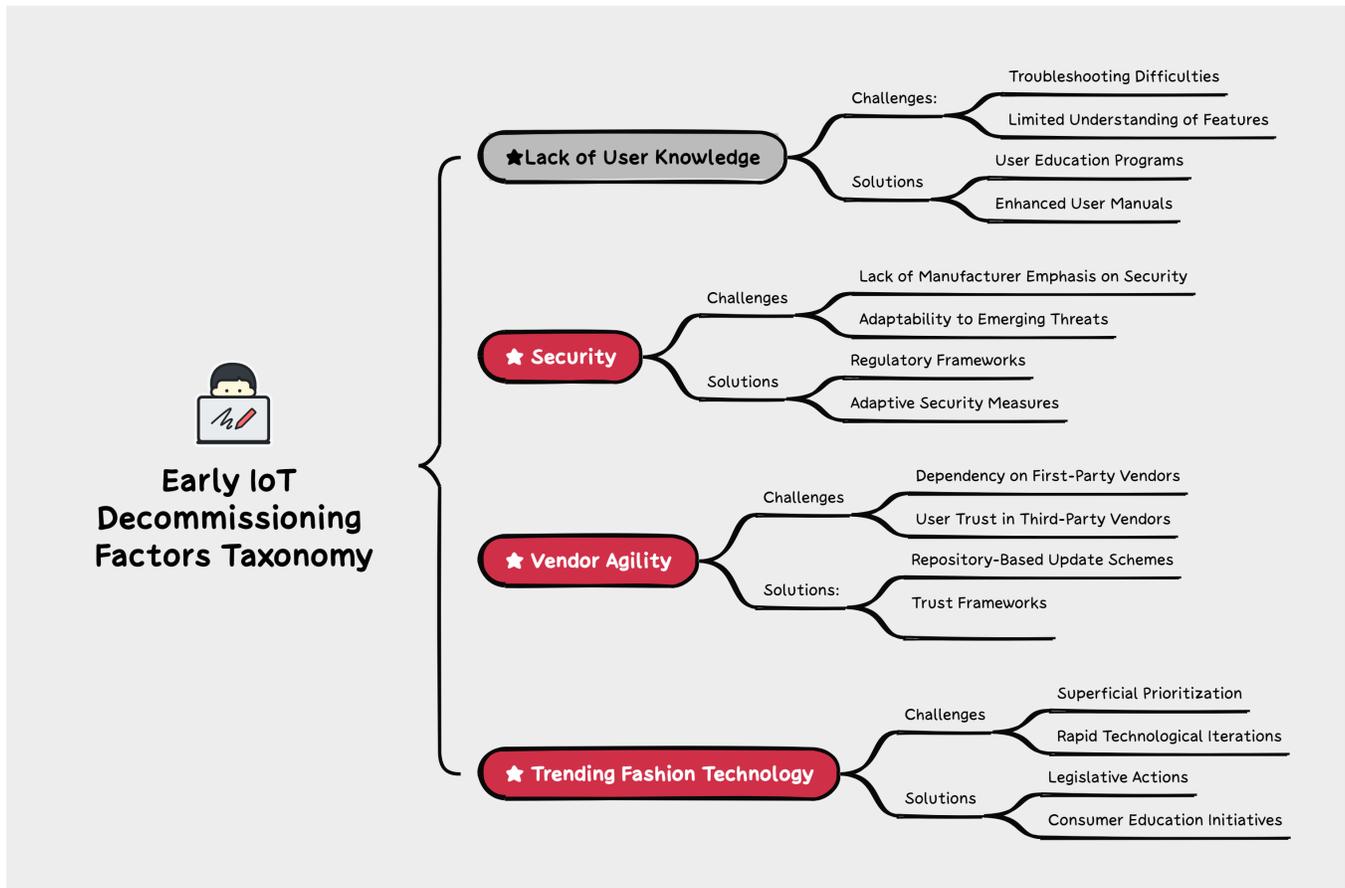

**Figure 1: Challenges and Solutions for Prolonging IoT Longevity Factors**

- Adaptive Security Measures: Developing adaptive security systems that can evolve alongside emerging threats is crucial for the long-term viability of IoT devices. Continuous research and development are essential in this context.

## 3.3 Vendor Agility

Vendors' agility is crucial in deciding the durability and sustainability of IoT devices. When a device's existing maintainer no longer supports it, transitioning to a different support channel is crucial. This shift guarantees that devices will keep receiving crucial upgrades, like as security patches, to reduce the chances of vulnerabilities and maintain the device's functioning and importance.

However, relying solely on the first-party vendor for software support poses significant challenges, particularly in scenarios where the vendor becomes unavailable or discontinues support for the device. In such cases, users are left with devices that are vulnerable to security threats and lacking critical updates[15], ultimately shortening the device's lifespan and contributing to electronic waste.

To address this challenge, an agile approach to vendor management is essential. By adopting repository-based update schemes with multiple stakeholders, IoT devices can transition seamlessly to alternative support channels when the first-party vendor is no longer available. This approach enables devices to continue receiving essential updates from trusted sources, thereby extending their software support period and enhancing their longevity[4].

However, while relying on third-party vendors for software support offers benefits in terms of extending the device's lifespan, users may hesitate to trust these entities with their devices. Concerns regarding data privacy, security, and reliability may deter users from entrusting third parties with responsibility for their devices.

As such, it is crucial to implement robust security measures and establish trust frameworks to ensure that users can confidently transition their devices to alternative support channels without compromising their privacy or security. Through careful consideration of these factors, the agility of vendors can be leveraged effectively to enhance IoT device longevity while addressing user concerns and ensuring a seamless transition process.

**Challenges:**

- Superficial Prioritization: Consumers may prioritize superficial qualities and design aesthetics over the functional longevity of devices, leading to premature replacements.
- Rapid Technological Iterations: The fast-paced release of similar devices with minor enhancements encourages consumers to upgrade frequently, contributing to electronic waste.

**Solutions:**

- Legislative Actions: Enacting legislation to discourage the unnecessary release of IoT devices with minimal upgrades can help curb the culture of constant technological turnover.
- Consumer Education Initiatives: Educating consumers about the long-term value and capabilities of their existing devices can shift the focus from trends to functionality, promoting a more sustainable approach to IoT usage.

### 3.4 Evolving Technological Trends

In today's rapid digital culture, IoT gadgets have evolved from functional tools to status symbols, similar to fashion items. Many businesses are introducing identical gadgets with little enhancements or cosmetic alterations, which attract people to constantly update their devices to keep up with the newest trends. This phenomenon reflects the dynamics of a fashion show, where newness and perceived social standing influence customer behavior, leading to a widespread culture of consumerism and frequent gadget upgrades.

Fashionable technological trends increase the problem of early disposal and play a major role in the growing e-waste dilemma. Consumers routinely abandon perfectly working IoT devices in favor of newer ones, contributing to resource depletion and environmental deterioration. Furthermore, the quick replacement of equipment not only depletes natural resources but also creates substantial difficulties in electronic waste handling and disposal.

To tackle the issue of trendy fashion technology, a comprehensive strategy involving legislative actions, consumer education, and industry responsibility is needed. Legislative actions may include enacting legislation to reduce the spread of unnecessary IoT devices.

**Challenges:**

- Dependency on First-Party Vendors: Relying solely on the first-party vendor for software support may pose challenges if the vendor becomes unavailable or discontinues support.
- User Trust in Third-Party Vendors: Users may hesitate to transition their devices to third-party vendors due to concerns about data privacy, security, and reliability.

**Solutions:**

- Repository-Based Update Schemes: Implementing repository-based update schemes involving multiple stakeholders can facilitate seamless transitions to alternative support channels when the first-party vendor is no longer available.
- Trust Frameworks: Establishing robust trust frameworks and security measures can instill confidence in users to transition their devices to third-party vendors, ensuring continued support without compromising security.

## 4 PROPOSED SOLUTIONS LITERATURE

### 4.1 Leveraging Blockchain for Enhancing IoT Longevity

Blockchain technology holds promise for improving the longevity of IoT devices by providing a decentralized and secure framework. Research by [16] outlines a paradigm for a trustworthy IoT ecosystem, leveraging blockchain's transparency and immutability to address the complexity and security challenges inherent in IoT networks. Their work highlights the potential of blockchain, particularly in securing IoT device provisioning, tracking, and decommissioning processes through smart contracts and decentralized identity management.

Furthermore, studies such as those by [8] and [21]support the efficacy of blockchain in enhancing IoT data security, integrity, and trust management. Additionally, the SerenIoT framework proposed by Kumar et al. extends this concept, offering a comprehensive blockchain-based solution for enhancing the security and efficiency of IoT ecosystems.

However, implementing blockchain algorithms in real-world IoT scenarios presents several challenges. The diverse categories of IoT devices and the potential for unlimited chains introduce scalability and resource consumption issues. Moreover, the latency and delays inherent in blockchain technology may not align well with the real-time requirements of IoT applications, particularly at scale. Addressing these challenges will be crucial for realizing the full potential of blockchain in enhancing the longevity and sustainability of IoT networks.

### 4.2 Sustainable Design For IoT

In the context of IoT, the strategic incorporation of sustainability into architectural design is crucial to guarantee the durability and sustainability as these devices become more widespread in everyday life. Requirements engineering serves as a critical framework for achieving this objective, offering a structured approach to identifying, prioritizing, and incorporating sustainability elements into the design process[14].

Requirements engineering plays a pivotal role in identifying key sustainability issues specific to IoT devices. By conducting thorough assessments of long-term system demands and considering environmental objectives alongside short-term goals, requirements engineers can anticipate and address potential sustainability challenges early in the design phase. This proactive approach ensures that sustainability considerations are integrated into the design from the outset, rather than being treated as an afterthought.

Moreover, requirements engineering facilitates the prioritization of sustainability concerns by engaging stakeholders and conducting trade-off analyses[2]. By involving stakeholders representing diverse perspectives and interests, requirements engineers can collaboratively identify and prioritize critical sustainability factors relevant to IoT device design. This inclusive approach fosters a shared understanding of sustainability goals, encouraging collective responsibility among stakeholders.

Further, the proposed exercise-mode logic [11] offers a proactive strategy to extend the lifespan of IoT devices within this sustainability-driven architectural design framework. By balancing the duty cycle of critical paths during idle periods, the exercise-mode logic systematically prevents wearout effects and enhances device reliability while minimizing hardware overhead. Integrated into the requirements engineering process, this logic ensures that sustainability considerations extend to device longevity, contributing further to the overall eco-friendliness of IoT systems.

Additionally, requirements engineering establishes a common knowledge base for sustainability within the IoT ecosystem. By providing stakeholders with a shared understanding of sustainability concepts and best practices, requirements engineers enable effective communication and coordination throughout the design process. This empowers stakeholders to align their efforts towards sustainable design outcomes, enhancing overall project success. Furthermore, requirements engineering facilitates the integration of sustainability considerations into IoT architectural decision-making. Through specialized viewpoints and methods, sustainability aspects are systematically analyzed and incorporated into designs, ensuring functional and environmentally responsible IoT systems[2].

However, the successful implementation of sustainable architectural designs for IoT devices requires intensive legislation to regulate manufacturing practices and ensure compliance with environmental standards. Additionally, these designs must be thoroughly documented to enable users to better understand their devices and take proactive actions to maintain their sustainability. By addressing these challenges and leveraging requirements engineering principles, the IoT industry can strive towards creating more durable, environmentally friendly devices that contribute to a sustainable future.

### 4.3 Revolution IoT Recycling

The conventional approach to recycling IoT devices often appears cost-prohibitive and unattractive due to the high costs involved, which can surpass the original purchase price of the devices. However, a paradigm shift in recycling methodology holds the potential for significant benefits. By implementing efficient collection and return systems for IoT devices, manufacturers can gain valuable insights into reusing components in future device iterations. This method not only encourages environmental sustainability but also offers economic benefits for businesses and customers[13]. Dedicated garbage bins for IoT devices need to be strategically positioned in cities and households to provide efficient collection and return systems. Legislative support and government aid are essential for creating and sustaining these infrastructural elements. Governments should encourage customers to return their outdated IoT devices to the original makers, which helps reintegrate discarded equipment into the manufacturing cycle. This closed-loop technology minimizes electrical waste and provides producers access to vital resources for future product development. Research in IoT recycling indicates that advanced methods like material recovery and component reintegration might improve the sustainability of recycling procedures[23]. Manufacturers may recover valuable materials from abandoned gadgets using innovative technology and procedures to reduce environmental impact and improve resource efficiency.

### 5 DISCUSSION

This research highlighted the significance of multidisciplinary collaboration in tackling the difficulties related to IoT device lifetime. By combining knowledge from several disciplines including technology, environmental science, policy-making, and consumer behavior, stakeholders may create comprehensive solutions that take into account the intricate relationships among technological, social, and environmental elements. This highlights the need of working together on research projects, forming collaborations with industries, and engaging in discussions across different sectors to encourage innovation and support sustainability in the IoT ecosystem.

Regulatory frameworks have a crucial role in influencing the sustainability of IoT devices. This paper recognizes the significance of legislative interventions in encouraging firms to focus on sustainability. However, further research is required to assess the efficacy of current rules and pinpoint areas that may be enhanced. This may entail comparing regulatory methods in other jurisdictions, assessing how industry standards influence product design and lifecycle management, and supporting legislative changes that encourage environmental stewardship and consumer protection.

The discussion also emphasizes the need of enabling consumers to make well-informed decisions regarding their IoT devices. Consumer education campaigns were suggested as a possible answer, but it is crucial to also take into account the wider socio-economic aspects that impact consumer behavior, including affordability, accessibility, and cultural norms. Future study might investigate methods to encourage digital literacy, improve transparency in product labeling and marketing, and promote community participation to bolster sustainable consumption patterns.

Lastly, this study highlights the importance of technical innovation in promoting sustainability in the IoT business. The research highlighted many technologies, including blockchain, edge computing, and artificial intelligence, that may improve gadget lifespan and lessen environmental harm. It is important to acknowledge the inherent constraints and trade-offs of these technologies, such as scaling challenges, data privacy concerns, and energy usage. Future research should concentrate on overcoming these problems and investigating innovative paths for technological advancement that emphasize both utility and sustainability.

### 6 CONCLUSION

This study has explored the intricate issues related to the longevity and sustainability of IoT devices in the current digital environment. The article has highlighted the significance of addressing causes leading to premature decommissioning and providing novel ways to enhance the resilience and eco-friendliness of the IoT ecosystem.

The research has identified many significant concerns such as user awareness gaps, security risks, vendor agility, and the impact of trend-driven customer behavior. The problems lead to early disposal of IoT devices and worsen the electronic waste problem, causing substantial environmental and economic issues. Nevertheless, within these constraints are potential for advancement and creativity. By adopting a comprehensive approach that integrates technology innovation, legislative interventions, consumer education, and industry responsibility, stakeholders in the IoT ecosystem can work towards enhancing device longevity and reducing environmental impact.

The proposed solutions literature has offered promising avenues for addressing these challenges. Leveraging blockchain technology holds potential for enhancing device provisioning, tracking, and decommissioning processes, while systematic methods of requirements engineering can integrate sustainability considerations into IoT architectural design from the outset. Additionally, adopting agile vendor management practices and reimagining recycling methodologies can further contribute to extending the lifespan of IoT devices and minimizing electronic waste.

It is crucial for stakeholders to work together across different fields and industries to properly put these solutions into action. Regulatory frameworks should be created to encourage firms to emphasize sustainability in their design and manufacturing processes. Consumer awareness efforts can enable people to make educated choices and engage in responsible consumption behaviors.

In conclusion, by embracing innovation, collaboration, and accountability, the IoT industry can pave the way for a more sustainable and resilient digital future. By addressing the challenges outlined in this study and implementing the proposed solutions, we can strive towards a world where IoT devices are not only technologically advanced but also environmentally sustainable, contributing to a more sustainable future for generations to come.